# Transport properties of the top and bottom surfaces in monolayer $MoS_2$ grown by chemical vapor deposition†

Sora Kurabayashi[1], and Kosuke Nagashio[*,1,2]

[1]Department of Materials Engineering, The University of Tokyo, Tokyo 113-8656, Japan
[2]PRESTO, Japan Science and Technology Agency (JST), Tokyo 113-8656, Japan
*nagashio@material.t.u-tokyo.ac.jp

The advantage of $MoS_2$, compared with graphene, is the direct growth on various oxide substrates by chemical vapor deposition (CVD) without utilizing catalytic metal substrates, which facilitates practical applications for electronics. The carrier mobility is, however, degraded from the intrinsic limit mainly due to short-range scattering caused by S vacancies formed during CVD growth. If the upper limit for the crystallinity of CVD-$MoS_2$ on oxide substrates is determined by the $MoS_2$/substrate interaction during growth, it will hinder the advantage. In this study, we investigated the interaction between monolayer $MoS_2$ and a $SiO_2$/Si substrate and the difference in crystallinity between the top and bottom S surfaces due to the $MoS_2$/substrate interaction. Raman and photoluminescence spectroscopy indicated that doping and strain were induced in $MoS_2$ from the substrate, but they could be removed by transferring $MoS_2$ to a new substrate using polymers. The newly developed polymer-transfer technique enabled selective transfer of the bottom or top surface of CVD-$MoS_2$ onto a new $SiO_2$/Si substrate. The metal-insulator transition was clearly observed for both the normal and inverse transfers, suggesting that the crystallinity of CVD-$MoS_2$ is high and that the crystallinity of the bottom surface interacting with the substrate was similar to that of the top free surface. These results provide positive prospects for the further improvement of the crystallinity of $MoS_2$ on oxide substrates by reconsidering the growth conditions.

## Introduction

$MoS_2$, a two-dimensional layered semiconductor, has been intensively investigated as a potential candidate for a high-performance transistor because its atomically thin layers are able to suppress short-channel effects.[1-3] Based on the large band gap of 1.8 eV for an $MoS_2$ monolayer, a field-effect transistor (FET) with a current on/off ratio of $\sim 10^8$ has been successfully demonstrated on a $SiO_2$/Si substrate.[4-11] The most important advantage of $MoS_2$, compared with graphene, is the direct growth of $MoS_2$ can be accomplished on various oxide substrates by chemical vapor deposition (CVD) and physical vapor deposition (PVD) without utilizing catalytic metal substrates,[12-17] which facilitates practical applications. In spite of successful FET operation, carrier mobilities attained by transport measurements at room temperature (RT)[18-20] are still far from the phonon-limited value of 410 $cm^2$/Vs predicted theoretically.[21] The major scattering factors for the mobility deterioration have been ascribed to the localized band-tail states caused by short-range disordered structural defects, such as S vacancies[18,22], and Coulomb traps.[6,23] From detailed observations by scanning transmission electron microscopy, the typical defect density has been reported to be $\sim 1 \times 10^{13}$ $cm^{-2}$.[18,22,24] The dominant defect type in mechanical exfoliation (ME) and CVD samples has been revealed to be S vacancies with one or two S atoms absent, while antisite defects with one Mo atom replacing one or two S atoms has been shown in PVD samples.[24] Therefore, the reduction of defects in $MoS_2$ is highly desired to further improve device performances.

When monolayer $MoS_2$ is grown by CVD on oxide substrates, such as $SiO_2$/Si,[25] sapphire,[17] and mica,[26] the photoluminescence (PL) peak intensity is generally higher than that for mechanically exfoliated $MoS_2$ on a $SiO_2$/Si substrate, and the full width at half maximum (FWHM) is smaller. These results suggest that CVD-$MoS_2$ has better crystallinity than ME-$MoS_2$. On the other hand, the mobility from transport measurements shows the opposite trend. This contradiction is expected to be attributed to the CVD-$MoS_2$/substrate interaction during growth. In first principles calculations for $MoS_2$ on $SiO_2$,[27] it is suggested that O-dangling bonds at the $SiO_2$ surface are chemically bonded with S atoms in $MoS_2$, which causes hole doping in $MoS_2$. This may occur similarly for all oxide substrates. Here, it is well known that both conduction and valence bands are composed of $d$-$d$ splitting in Mo due to the ligand field for trigonal prism coordination.[22,28] Therefore, the electron in the conduction band is mainly located in Mo atoms. The electron density will be lopsided to the interface between $MoS_2$ and gate insulator when the gate bias is sufficiently applied. Therefore, the transport properties of $MoS_2$ might be largely affected by interactions with the oxide substrates, and the number of S vacancies on the bottom surface of monolayer $MoS_2$ on $SiO_2$ with inhomogeneous charges[29] is expected to be larger than that on the top free surface. It is, however, difficult to clarify the number of S vacancies for the top and bottom surfaces by scanning transmission electron microscopy (STEM) due to the same contrast. More importantly, there is an unfavorable possibility that the upper limit for the crystallinity of CVD-$MoS_2$ on oxide substrates might be determined by the $MoS_2$/substrate interaction during growth, which will hinder the advantage of the direct growth of $MoS_2$ on oxide substrates.

In this study, the interaction between CVD-$MoS_2$ and a $SiO_2$/Si substrate was systematically evaluated using Raman and PL measurements by transferring as-grown $MoS_2$ to a new $SiO_2$/Si substrate. Moreover, the newly developed polymer-transfer technique enabled to selectively transfer the bottom or top surfaces of CVD-$MoS_2$ on a new $SiO_2$/Si substrate, and the transport properties of the bottom and top surfaces of monolayer $MoS_2$ were measured using back-gate modulation to qualitatively clarify the crystallinity of the bottom and top surfaces.

**CVD growth and normal transfer process**

**Fig. 1a** schematically illustrates the experimental setup for CVD growth of $MoS_2$ on a $SiO_2/Si$ substrate. The S (150 mg) and $MoO_3$ (25 mg) powders in quartz boats were separately placed in a quartz reaction tube. The surface treatment of the $SiO_2/Si$ wafer is important to promote the nucleation of $MoS_2$.[30] In this study, to get reproducible results, the $SiO_2$ surface was slightly etched by a diluted HF solution (HF:deionized water (DIW) = 1:20) and followed by a DIW rinse. The $SiO_2/Si$ wafers were mounted on two different positions: (i) on top of the quartz boat of the $MoO_3$ powder and (ii) on a quartz substrate holder to adjust the height. It should be noted that either position was used in each growth run. During the synthesis, the $MoO_3$ powder was first heated to 700 °C under a nitrogen gas flow of 260 ml/min, then the S powder was subsequently heated to 240 °C because $MoO_3$ vapor is drastically suppressed when the $MoO_3$ surface is sulfurized,[30] as shown in **Supplementary Fig. S1**. As shown in **Fig. 1b,c**, single-crystal grains in the shape of well-defined equilateral triangles were widely distributed throughout the $SiO_2/Si$ wafer at the substrate position (ii), which is more suitable to fabricate single-crystal grain $MoS_2$ FET devices. For further characterization and device fabrication, position (ii) was selected.

The quality of the as-grown monolayer $MoS_2$ film was studied by PL and Raman spectroscopy and compared with that of ME-monolayer $MoS_2$. As shown in **Fig. 2a,b**, the PL intensity for the as-grown $MoS_2$ film is much larger than that for ME-$MoS_2$, and the FWHM of the PL peak for the as-grown $MoS_2$ is smaller than that for ME-$MoS_2$. These results are consistent with previous reports.[25] On the other hand, as shown in **Fig. 2c,d**, the Raman $E_{2g}$ peak for the as-grown $MoS_2$ is redshifted from that for ME-$MoS_2$, indicating that the in-plane vibration is mainly altered. Based on a previous uniaxial-strain experiment,[32] a strain of ~1 % is expected to be induced in the as-grown $MoS_2$ due to the interaction with the $SiO_2/Si$ substrate.

To reduce the interaction between $MoS_2$ and the substrate, $MoS_2$ grains grown on the $SiO_2/Si$ substrate were transferred to a new $SiO_2/Si$ substrate using a conventional polymer-transfer method with polymethylmethacrylate (PMMA) and polydimethylsiloxane (PDMS),[25,33] as schematically shown in **Fig. 3a**. After the transfer of $MoS_2$, PMMA residue was carefully removed by baking under $H_2/Ar$ gas flow for 1 hour at 200 °C. In this case, the bottom surface of the as-grown $MoS_2$ is attached to the new $SiO_2/Si$ substrate, which is called "normal transfer", as shown in **Supplementary Fig. S2a**. Interestingly, both the PL intensity and the FWHM of the PL peak for the normal-transfer $MoS_2$ were drastically reduced to the same level as those of ME-$MoS_2$, as shown in **Fig. 2a,b**. Moreover, the Raman $E_{2g}$ peak position also became comparable with that for ME-$MoS_2$. Based on this transfer experiment, it is evident that the strain induced during the CVD growth was released. The strong PL intensity observed for the as-grown $MoS_2$ does not indicate that the crystallinity of CVD-$MoS_2$ is better than that of ME-$MoS_2$. It is likely that the p-type doping due to the interaction between CVD-$MoS_2$ and the $SiO_2/Si$ substrate reduces the recombination of negatively charged trions and enhances the recombination of excitons.[34] Although the p-type doping is one of possibilities due to the slight blue shift of the $A_{1g}$ peak from ME-$MoS_2$ to the as-grown $MoS_2$,[35] it is not evident because $A_{1g}$ position for the normal transfer is comparable with that for as-grown $MoS_2$. Here, if the interaction between CVD-$MoS_2$ and the $SiO_2/Si$ substrate is characteristic of chemisorption rather than physisorption, it will be expected that the amount of defects, such as S vacancies, for the bottom surface of monolayer $MoS_2$ will be greater than that for the top free surface due to the change in the electronic structure around the chemical bonds.

The difference in the number of defects for the bottom and top surfaces can be qualitatively evaluated by comparing the transport properties for top and bottom gate modulation in the dual-gated devices beca use the carrier distribution along the thickness direction in $MoS_2$ is lopsided by the electric field effect, and electrons will be trapped at localized defect states below the conduction band induced by S vacancies.[22]

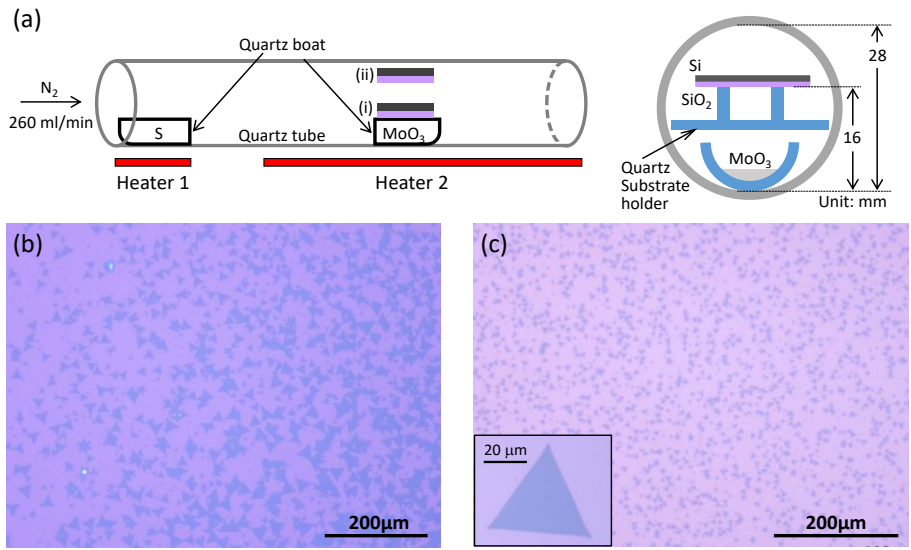

**Fig. 1**  (a) Schematic illustration of the $MoS_2$ CVD system. (b,c) Optical images of $MoS_2$ grown at positions (i) and (ii), respectively.

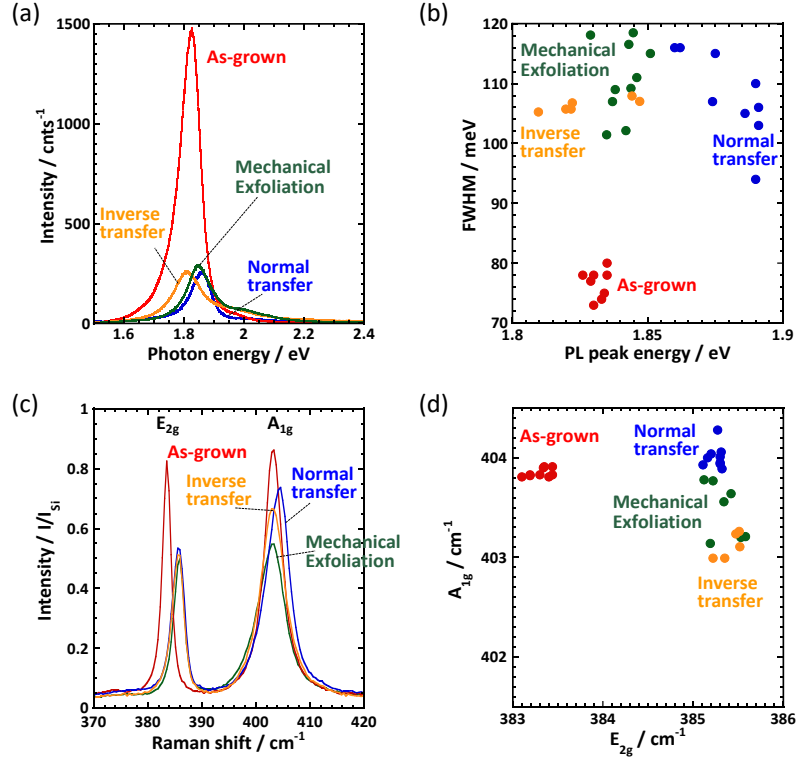

**Fig. 2** (a) PL spectra of the as-grown, ME, normal-transfer and inverse-transfer samples. (b) FWHM as a function of PL peak intensity for the four kinds of samples in (a). (c) Raman spectra of the as-grown, ME, normal-transfer and inverse-transfer samples. The peak intensities are normalized by the Si peak intensity at 520 cm$^{-1}$. (d) Relationship between $A_{1g}$ and $E_{2g}$ for the four kinds of samples in (c).

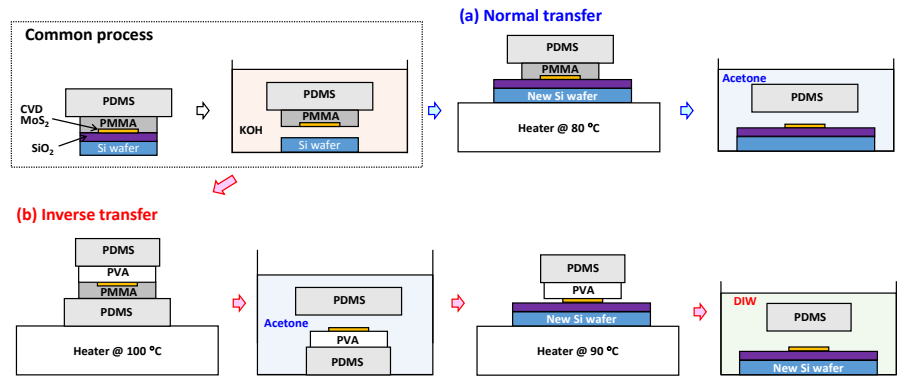

**Fig. 3** Polymer-transfer method used in this study. For the etching of SiO$_2$, MoS$_2$ coated with PMMA/PDMS was immersed in a diluted KOH solution (KOH:DIW = 3:2) overnight. (a) Normal transfer. In general, we waited for one day to dissolve PMMA in acetone. (b) Inverse transfer. MoS$_2$ sandwiched by PVA and PMMA was immersed in DIW for a few days to dissolve PVA. For both transfer methods, Ar/H$_2$ annealing was carried out at 200 °C for 1 hours to remove the resist residue and to improve the adhesion between MoS$_2$ and the new SiO$_2$/Si wafer.

## Dual-gate modulation

The fabrication process for the dual-gate FET devices is shown in **Fig. 4**. After the transfer of CVD-MoS$_2$ to a new SiO$_2$/Si substrate, a triangle shape was patterned into a multi-terminal device structure by conventional electron-beam (EB) lithography and CF$_4$/O$_2$ plasma etching, followed by the formation of source and drain electrodes (Ti/Au or Ni/Au). Subsequently, Y metal with the thickness of ~1 nm was deposited as a buffer layer via thermal evaporation at a rate of ~0.1 Å/s in an Ar atmosphere of 10$^{-1}$ Pa.[36] When Y metal was oxidized via O$_2$ annealing at 200 °C for 10 min, the two-probe conductivity was drastically reduced for the device with Ti/Au electordes, as shown in **Supplementary Fig. S3a**. This could be due to the oxidation of Ti electrode, since this degradation is quite limited for the two-probe device with Ni/Au electrodes. It should be noted that Ni/Au electrodes are mainly used, especially, for the low temperature measurements. Therefore, Y metal naturally oxidized in the laboratory atmosphere for 1 day. Al$_2$O$_3$ with a thickness of ~30 nm was deposited by atomic layer deposition (ALD).[37] The Al top-gate electrode was formed via EB lithography. The optical image for the final dual-gate FET is shown in **Fig. 4c**. The transport measurements were performed in a

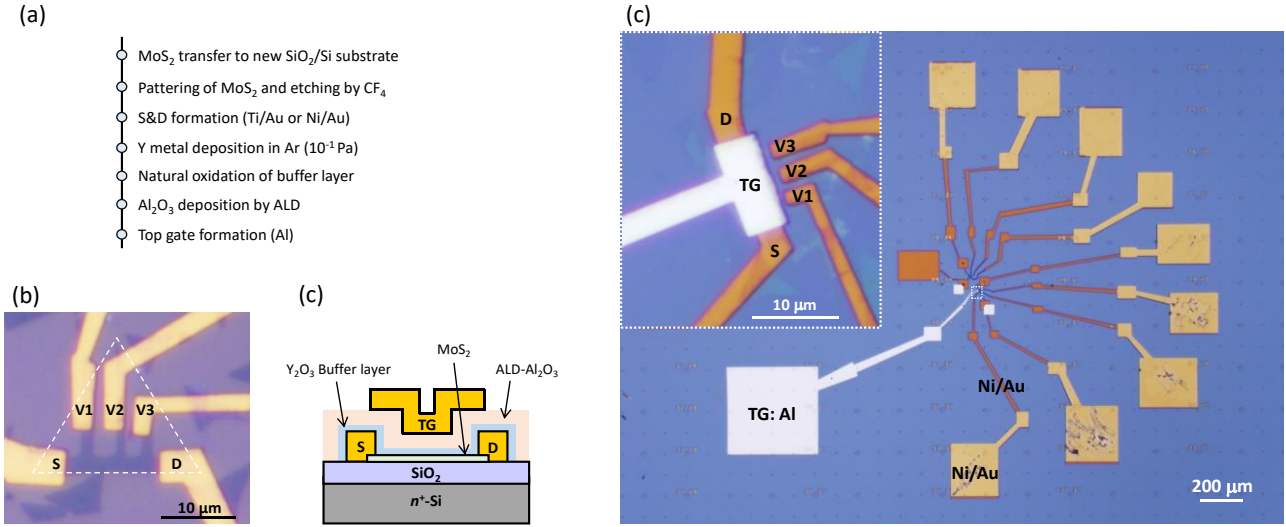

**Fig. 4** (a) Process flow for the top-gate MoS$_2$ FET. (b) Optical image of the back-gate MoS$_2$ FET with the multi-terminal structure after the shaping by CF$_4$ etching. The initial triangle shape of MoS$_2$ is shown by the white broken line. (c) Schematic illustration of cross-sectional view of the top-gate MoS$_2$ FET. (d) Optical image of typical dual-gate MoS$_2$ FET. Inset shows the magnified image of the device.

vacuum prober at different temperatures. **Fig. 5a** shows the four-probe conductivity as a function of the top-gate voltage ($V_{TG}$) at different back-gate voltages ($V_{BG}$) for a typical dual-gate device at room temperature (RT). The threshold voltage ($V_{th}$) continuously shifted with $V_{BG}$, which was controlled by the relative ratio of the capacitive coupling between the top and back gates with MoS$_2$. The trace of $V_{th}$ observed for the $V_{TG}$ sweep was plotted as a function of $V_{BG}$, as shown in **Supplementary Fig. S3b**. The slope ($S$) corresponds to $-C_{BG}/C_{TG}$, where $C_{BG}$ and $C_{TG}$ are the capacitances for the back gate and top gate, and $C_{BG}$ was 0.0384 μF/cm$^2$ for the 90-nm thick SiO$_2$ using $k_{SiO2} = 3.9$. Therefore, $C_{TG}$ was calculated to be 0.219 μF/cm$^2$, and the dielectric constant for Al$_2$O$_3$ with a Y$_2$O$_3$ buffer layer was

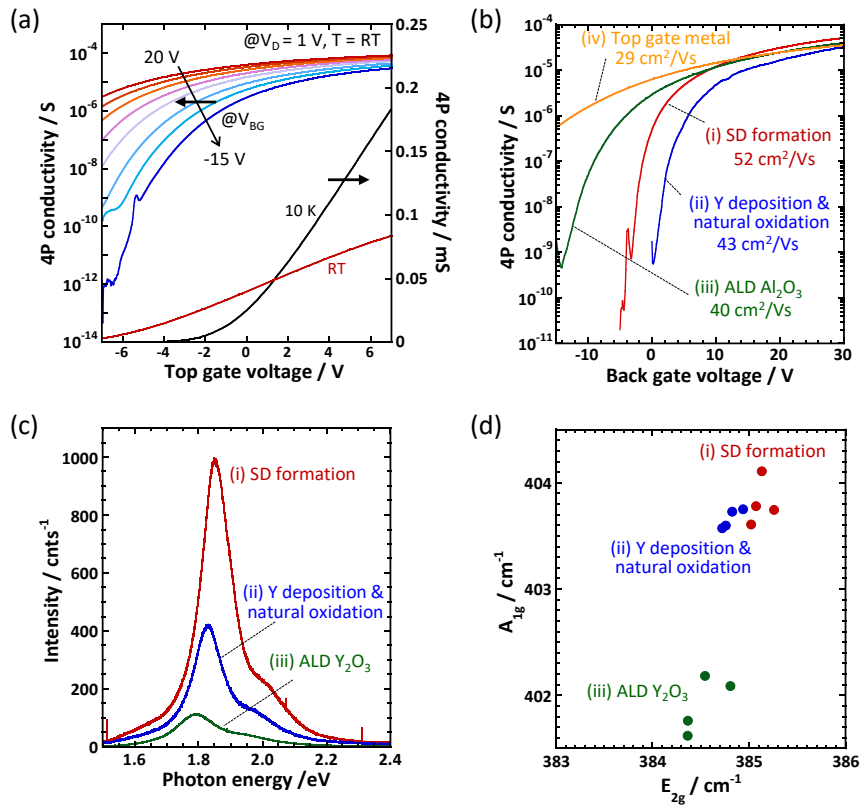

**Fig. 5** (a) Conductivity as a function of $V_{TG}$ at different $V_{BG}$ for the dual-gate device obtained by four-probe measurements. The left axis is in a log scale, while the right axis is in a linear scale. Conductivities as a function of $V_{TG}$ at $V_{BG}$ = 20 V at 10 K and RT are shown in the linear scale. (b) Conductivity as a function of $V_{TG}$ at $V_{BG}$ = 0 V obtained during each top-gate formation process. These data are also obtained by the four-probe measurements. (c) PL spectra of the as-grown, ME, normal-transfer and inverse-transfer samples. (d) Relationship between the A$_{1g}$ and E$_{2g}$ modes obtained during each top-gate formation process.

estimated to be ~7.7 using a thickness of ~31 nm. This is a reasonable value compared with that of typical ALD Al$_2$O$_3$ on graphene.[38] Using $C_{TG}$ and the transconductance, the field-effect mobility ($\mu_{FE}$) was calculated to be 32 cm$^2$/Vs at RT and 109 cm$^2$/Vs at 10 K, which are comparable with other literature data. It should be noted that $\mu_{FE}$ is slightly underestimated because the contribution of quantum capacitance is neglected in this calculation.[39] **Supplementary Fig. S3** summarizes previously reported device performances for top-gated CVD-grown monolayer MoS$_2$ FETs.[40-50]

Although a relatively high $\mu_{FE}$ was obtained for the top-gate modulation, it is not valid to compare the transport properties for the top- and back-gate modulation unfortunately. The four-probe conductivity as a function of $V_{BG}$ at different device fabrication process stages is shown in **Fig. 5b**. The $\mu_{FE}$ estimated for $V_{BG}$ modulation continuously decreased at each top-gate fabrication stage from 52 cm$^2$/Vs to 29 cm$^2$/Vs. This is solely due to the gate-stack formation because the effect of the contact electrodes was excluded by the four-probe measurement. The negative shift of $V_{th}$ indicates electron doping in the MoS$_2$ channel. According to previous reports, Raman A$_{1g}$ peak redshifts due to n-doping[35], and n-doping causes a transition from the recombination of excitons to the recombination of negatively charged trions in the PL spectra.[34] These are all consistent with each other, as can be clearly seen in **Fig. 5c,d**, where the Raman and PL spectra were obtained through the gate oxides. It should be noted that the strain introduced by the oxide deposition on MoS$_2$ is much smaller than that induced by the MoS$_2$/SiO$_2$ interaction during the CVD growth, which is confirmed by the comparison of **Fig. 2d** and **Fig. 5d**. However, it is not valid to qualitatively compare the difference in crystallinity for the bottom and top surfaces from the transport properties obtained by the top- and bottom-gate modulation because the transport properties for the back-gate modulation are drastically degraded due to the deposition of the top-gate oxides.

**Comparison of normal & inverse transfer**

To evaluate the transport properties of the top surface of the as-grown MoS$_2$, an "inverse transfer" method was newly developed, as shown in **Fig. 3b**. The key point was to use two different polymers: one is PMMA, which is dissolved in acetone, and the other is polyvinyl alcohol (PVA), which is dissolved in water. After the as-grown MoS$_2$ was detached from the old SiO$_2$/Si wafer using the PMMA/PDMS polymers, it was further sandwiched by the PVA/PDMS polymers. The sandwiched polymers with MoS$_2$ were immersed in acetone to dissolve PMMA, which enabled the inverse transfer to a new SiO$_2$/Si wafer, as shown in **Supplementary Fig. S2**. In this case, the transport properties of the top surface of the as-grown MoS$_2$ can be evaluated using the back-gate modulation without any deposition of oxides onto the MoS$_2$ channel.

After the inverse transfer, as expected, both the PL intensity and FWHM of the PL peak were drastically reduced to the same level as those of ME-MoS$_2$, as shown in **Fig. 2b,d**. Moreover, the Raman E$_{2g}$ peak again moved back to a similar position as that of ME-MoS$_2$, indicating the release of strain. The main difference between the normal and inverse transfers is that the Raman A$_{1g}$ peak for the inverse transfer shifted negatively in reference to ME-MoS$_2$, while it shifted positively for normal transfer. This suggests slight p-doping in MoS$_2$ for the inverse transfer, instead of n-doping for the normal transfer. Most importantly, no noticeable defect formation occurred during the polymer transfer, as the Raman intensity did not decrease, as shown in **Fig. 2c**.

Thus far, strong electron-electron correlation has been proposed to explain the metal-insulator transition (MIT) in MoS$_2$, which provides a universal threshold density $n_{MIT}$ ~ 10$^{13}$ cm$^{-2}$.[47] On the other hand, it has been reported that $n_{MIT}$ is dependent on the sample quality of ME-MoS$_2$, i.e., the trap-state density ($N_{tr}$), which is related to the number of S vacancies.[18] The MIT was clearly observed when $N_{tr}$ is smaller than ~5.5 × 10$^{12}$ cm$^{-2}$ after the repair of S vacancies, while it was not observed for larger $N_{tr}$ before the repair. These reported data are plotted as orange circles and as orange cross in **Fig. 6c**, where the circles and cross indicate the observation of the MIT and no observation of the MIT, respectively. Other literature data are also added in the figure and in **Supplementary Fig. S4**. Although the origin for the MIT is out of the focus of this paper, **Fig. 6c** suggests that the low-mobility samples do not show the MIT. Therefore, it may be possible to qualitatively evaluate the crystallinity for the bottom and top surfaces from the observation of the MIT.

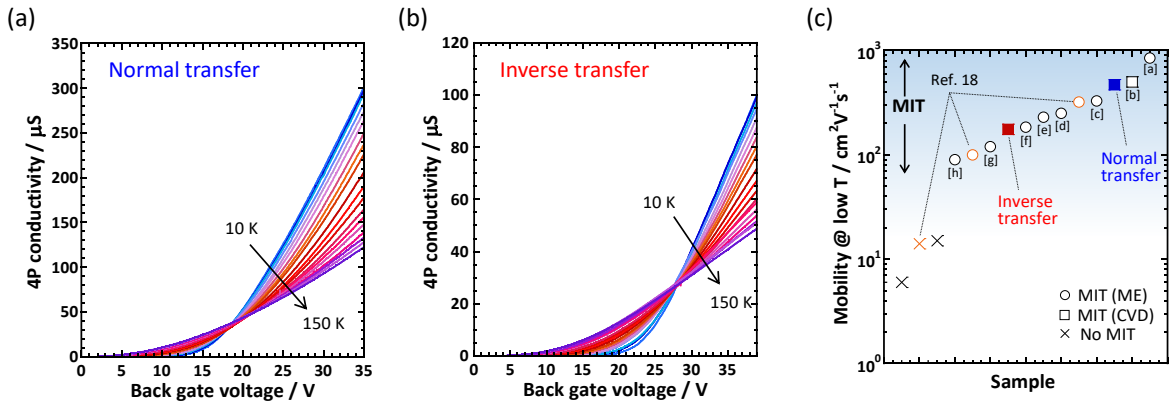

**Fig. 6** Conductivity as a function of $V_{BG}$ at different temperatures for the normal transfer (a) and inverse transfer (b). These data are obtained by the four-probe measurements. (c) Mobility obtained at low temperature (typically ~4 - 20 K) for different samples. The references are shown in **Supplementary Fig. S4**.

**Fig. 6a,b** show the four-probe conductivities as a function of $V_{BG}$ at different temperatures for normal- and inverse-transfer devices. The MITs are clearly observed at ~40 μS for normal transfer and at ~20 μS for inverse transfer, as also shown in **Supplementary Fig. S4**. According to the Matthiessen's rule, the total mobility can be described as the sum of three scattering contributions, i.e., temperature-dependent phonon scattering, short-range defect scattering, and long-range Coulomb impurity scattering, as follows,

$$\frac{1}{\mu_{total}(T)} = \frac{1}{\mu_{PH}(T)} + \frac{1}{\mu_{SR}} + \frac{1}{\mu_{LR}}.$$

To gain insight into the scattering mechanism, $\mu_{FE}$ is plotted as a function of temperature, as shown in **Fig. 7a**. In the temperature range of 60 - 300 K, phonon-limited behavior of $\mu \sim T^{-\gamma}$ is observed with the exponent γ equal to approximately 1.2 and 1.0 for the normal and inverse transfer, respectively. Further decreasing the temperature to 10 K results in the mobility becoming constant at 480 cm$^2$/Vs and 170 cm$^2$/Vs for the normal and inverse transfer, respectively, due to the short-range defect scattering or long-range Coulomb impurity scattering. These behaviors are consistent with previous experiments[18,25,47-52] and theoretical simulations.[21] Here, the mobility at 10 K is saturated with increasing carrier density ($V_{BG}$), as shown in **Fig. 7b**. This suggests that the scattering due to long-range Coulomb impurities can be ruled out for both the normal and inverse transfer, because increasing carrier density generally enhances the screening of the Coulomb potential.[9,25] Therefore, the different interface conditions due to different polymers used in the normal and inverse transfer are not dominant issue, because the present CVD MoS$_2$ is controlled by the short-range defect scattering.

For both the normal and inverse transfer, the MIT was observed, and their low-temperature mobilities were in the range of previously reported MIT observations, mainly for ME-MoS$_2$. Although it was expected that the inverse transfer would provide superior transport properties to that of the normal transfer due to the absence of an interaction with the SiO$_2$/Si substrate during the CVD growth, this was not the case. Nevertheless, the present crystallinity of CVD-MoS$_2$ is high enough to observe the MIT, suggesting the crystallinity for CVD-MoS$_2$ is comparable with that for bulk MoS$_2$ crystals grown naturally without any interaction with the substrate. The difference in the crystallinity for the top and bottom surfaces is negligible. Moreover, it is reported that the nucleation of MoS$_2$ on SiO$_2$ is difficult and that MoS$_2$ starts to grow from nominally oxi-chalcogenide nanoparticles as heterogeneous nucleation sites.[53] Based on this discussion, the MoS$_2$/SiO$_2$ interaction could be physisorption, which suggests positive prospects for further improvement of the crystallinity of MoS$_2$ on oxide substrates by reconsidering the growth conditions.

## Conclusions

The transport properties of the top and bottom surfaces of monolayer CVD-MoS$_2$ were investigated to elucidate the difference in crystallinity for the top and bottom surfaces due to the interaction between MoS$_2$ and the SiO$_2$/Si substrate during CVD growth. Raman and PL spectroscopy indicated that doping and strain were induced in MoS$_2$ from the substrate, but they could be removed by transferring MoS$_2$ to new substrate using polymers. The transport properties for both the normal and inverse transfers suggested that the present crystallinity of CVD-MoS$_2$ is high enough to observe the MIT and that the difference in crystallinity for the top and bottom surfaces is negligible. These results suggest positive prospects for the further improvement of the crystallinity of MoS$_2$ on oxide substrates by reconsidering the growth conditions.

**Acknowledgements**

This research was partly supported by the JSPS Core-to-Core Program, A. Advanced Research Networks, JSPS KAKENHI Grant Numbers JP25107004, JP16H04343, JP16K14446, & JP26886003, JST PRESTO Grant Number JPMJPR1425, Japan.

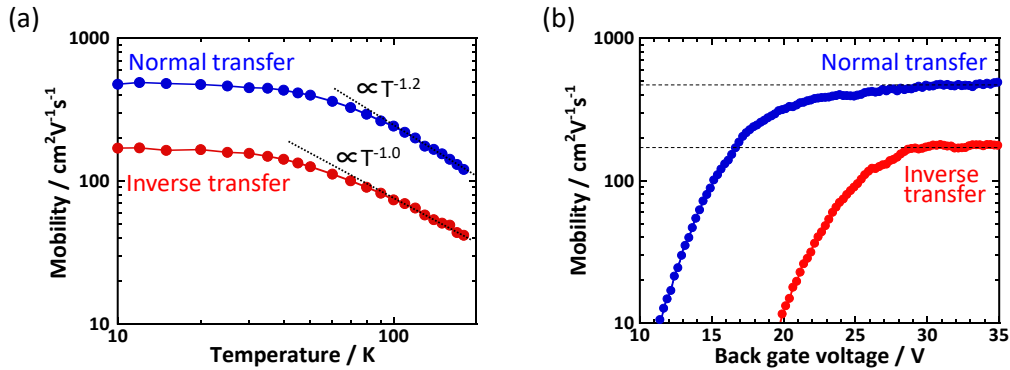

**Fig. 7** (a) Field-effect mobility as a function of temperature for the normal and inverse transfers. (b) Mobility as a function of back gate voltage measured at 10 K.

# Transport properties of the top and bottom surfaces in monolayer MoS$_2$ grown by chemical vapor deposition

*Sora Kurabayashi[1], and Kosuke Nagashio*[*,1,2]

[1]Department of Materials Engineering, The University of Tokyo, Tokyo 113-8656, Japan
[2]PRESTO, Japan Science and Technology Agency (JST), Tokyo 113-8656, Japan
*nagashio@material.t.u-tokyo.ac.jp

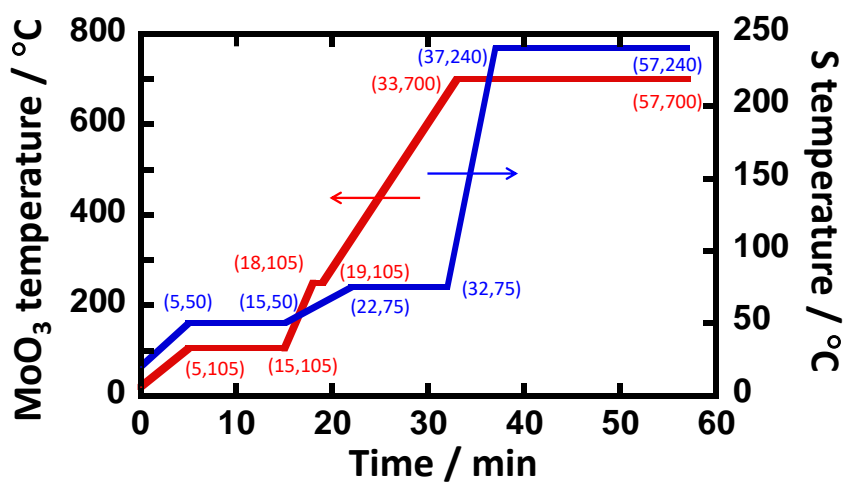

**Figure S1:** The program set up for the temperatures of MoO$_3$ and S powders.

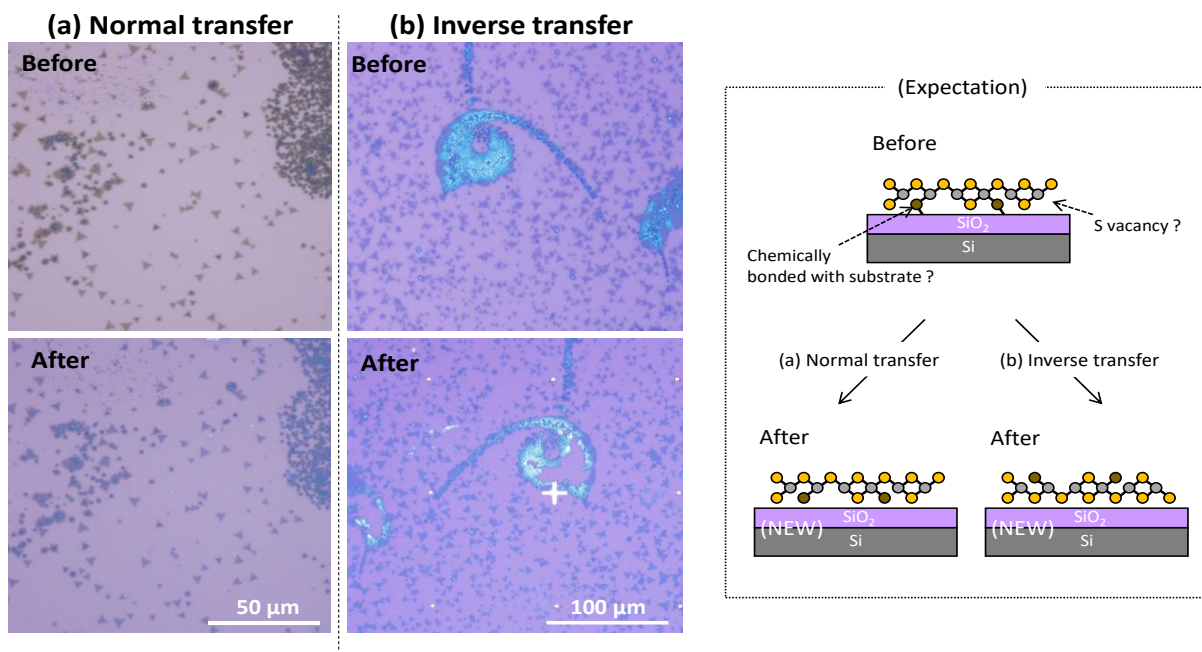

**Figure S2:** Optical images before and after the transfer of as-grown $MoS_2$ to new $SiO_2$/Si substrate. (a) normal transfer, and (b) inverse transfer. Expected schematic illustration for normal and inverse transfer of as-grown $MoS_2$ is also shown in the right figure. However, the transport properties for both normal and inverse transfers in **Fig. 6** in the main text suggest that the present crystallinity of CVD-$MoS_2$ is high enough to observe MIT and the difference in the crystallinity for top and bottom surfaces could be negligible.

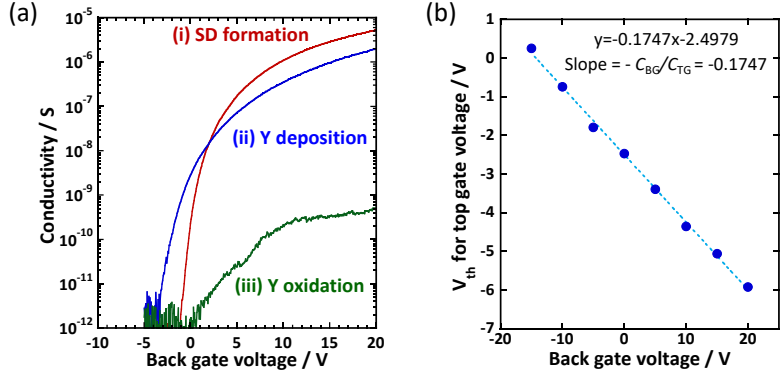

Table: Literature data for top-gate CVD-monolayer MoS$_2$ FET devices.

| Year | Method | Material | Thickness nm | Mobility cm$^2$/Vs | Contact resistance | Ref. |
| --- | --- | --- | --- | --- | --- | --- |
| 2013 | ALD | Al$_2$O$_3$ | 16 | 21.6 | excluded | Nano lett. 2013, 13, 2640. |
| 2014 | ALD | HfO$_2$ | 30 | 7 | excluded | Nature comm. 2014, 5, 3087. |
| 2015 | ALD | Al$_2$O$_3$ | 25 | 24 | Included | Appl. Phys. Lett. 2015, 106, 062101. |
| 2015 | ALD | HfO$_2$ | 30 | 30 | excluded | Nature 2015, 520, 656. |
| 2015 | ALD | HfO$_2$ | 30 | 63 | excluded | Nano lett. 2015, 15, 5039. |
| 2016 | ALD | HfO$_2$ | 30 | 42.3 | included | Appl. Phys. Lett. 2016, 203105. |
| 2016 | ALD | HfO$_2$ | 30 | 54 | excluded | Adv. Mater. 2016, 28, 1818. |
| Present | ALD | Al$_2$O$_3$ | 31 | 32 | excluded | |

**Figure S3:** (a) Two-probe conductivity as a function of $V_{TG}$ at $V_{BG} = 0$ V obtained during the each top gate formation process for the device with Ti/Au electrodes. It is clear that the conductivity was drastically reduced after the oxidation of the Y metal buffer layer at 200 °C for 10 min. (b) The trace of $V_{th}$ for the $V_{TG}$ sweep in **Fig. 5a** in the main text is plotted as a function of $V_{BG}$. The linear relation can be seen. Table shows the literature data for top-gated CVD-grown monolayer MoS$_2$ FET.

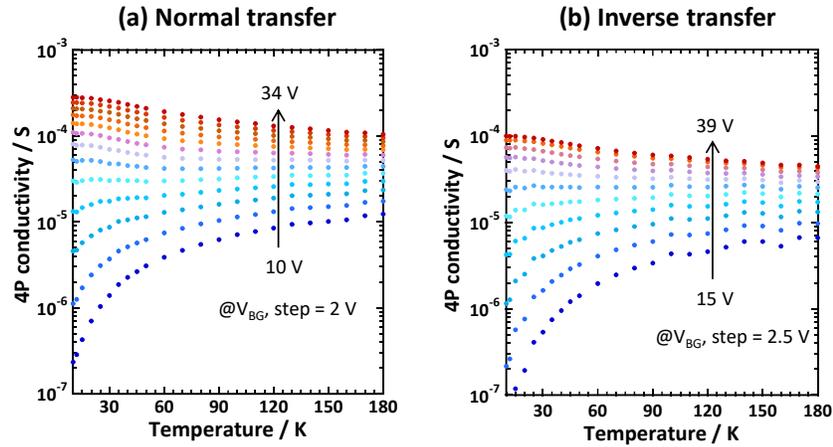

Table: Literature data for monolayer MoS$_2$ device showing MIT.

| No. | Year | Sample | Substrate | Gate | Mobility cm$^2$/Vs | Ref. |
|---|---|---|---|---|---|---|
| a | 2016 | ME | HfO$_2$ | BG | 847 | Adv. Mater. 2016, 28, 547. |
| b | 2014 | CVD | SiO$_2$ | BG | 500 | Nano Lett. 2014, 14, 1909. |
|   | Present | CVD | SiO$_2$ | BG | 470, 175 | Normal & Inverse transfers |
| c | 2015 | ME | h-BN | BG | 328 | Nano Lett. 2015, 15, 3030. |
|   | 2014 | ME | SiO$_2$ | BG | 320, 100 | Nature comm. 2014, 5, 5290. |
| d | 2013 | ME | SiO$_2$ | BG | 250 | Nano Lett., 2013, 13, 4212. |
| e | 2014 | ME | ion gate/SiO$_2$ | TG | 230 | Sci. Rep. 2014, 4, 7293. |
| f | 2013 | ME | HfO$_2$ | TG | 184 | Nature mater. 2013, 12, 815. |
| g | 2103 | ME | SiO$_2$ | BG | 120 | Appl. Phys. Lett., 2013, 102, 173107. |
| h | 2015 | ME | h-BN | TG | 90 | Nature comm. 2015, 6, 6088. |

**Figure S4:** Four probe conductivity as a function of temperatures for (a) normal transfer and (b) inverse transfer. Table shows the literature data for monolayer MoS$_2$ device showing MIT.